\documentstyle [11pt,paspconf,epsf]{article}

\begin{document}

\title{
Baryonic Density from Primordial Li}
\author{
Piercarlo Bonifacio }
\affil{
Osservatorio Astronomico di Trieste, via G.B. Tiepolo 11, 34131, Trieste Italy}

\author{Paolo Molaro}
\affil{
Osservatorio Astronomico di Trieste, via G.B. Tiepolo 11, 34131, Trieste Italy}
\begin{abstract}
Lithium abundances in a 
selected sample of halo stars have been 
revised by using  new accurate T$_{\rm eff}~$. 
From 41  plateau stars we found no 
evidence for intrinsic dispersion, 
a tiny trend with T$_{\rm eff}~$ and no  trend with  [Fe/H].
These results argue against depletion by either stellar winds,
diffusion or rotational mixing. Therefore the Li observed in PopII
stars provides a reliable estimate of the baryonic density.
A more detailed discussion can be found
in Bonifacio \& Molaro (1996).
\end{abstract}


\section{ Introduction}

The primordial nature of the Li observed in Pop II stars
rests on the existence of the Spite plateau, i.e. 
constant Li abundance for dwarfs with ${T}_{\rm eff}~$ $>$5700 K and
 [Fe/H]$<$-1.5. 
Recently, some authors  have claimed the existence
of trends of Li abundance both with ${T}_{\rm eff}~$ and  [Fe/H],
and intrinsic dispersion on the plateau
(Deliyannis et al 1993, Norris et al 1994, Thorburn 1994, Ryan et al 1996). 
In addition,
one star in M92 has been found with  [Li]=2.5, which is well above
the plateau value (Deliyannis et al 1995). These results would
argue in favour of a depletion of Li in the atmospheres
of Pop II stars, weakening its cosmological 
significance.
The existence of the Spite plateau
has been defended by   Molaro et al (1995) and Spite et al (1996).

\section{ A Selected 
 Sample for Lithium}

 ${T}_{\rm eff}~$ is the major source
of error in the Li abundance determination
and a precise determination of ${T}_{\rm eff}~$ is  necessary
in order to discuss possible trends or dispersion
on the plateau.
The best method to determine ${T}_{\rm eff}~$, short of a
direct measure of the angular diameter, is the IRFM
(Blackwell et al 1990).
Here we revise the Li abundance for 64 Pop II stars,
which are about 2/3 
of the Pop II stars observed for Li,  using
new IRFM ${T}_{\rm eff}~$  obtained by Alonso et al (1996). 
Gravities have been recomputed for all the stars
and results are shown in Fig. 1.
 The exclusion of the few possible subgiants from the
 sample does not alter significantly the results presented in this
 paper.
The Li EWs have been taken from the literature. For multiple 
measurements we adopted the weighted average. 
The errors have been taken from the 
original papers, when available or estimated according the Ryan et al (1996)
prescriptions.

\subsection{Models }

\begin{figure}
\plotone{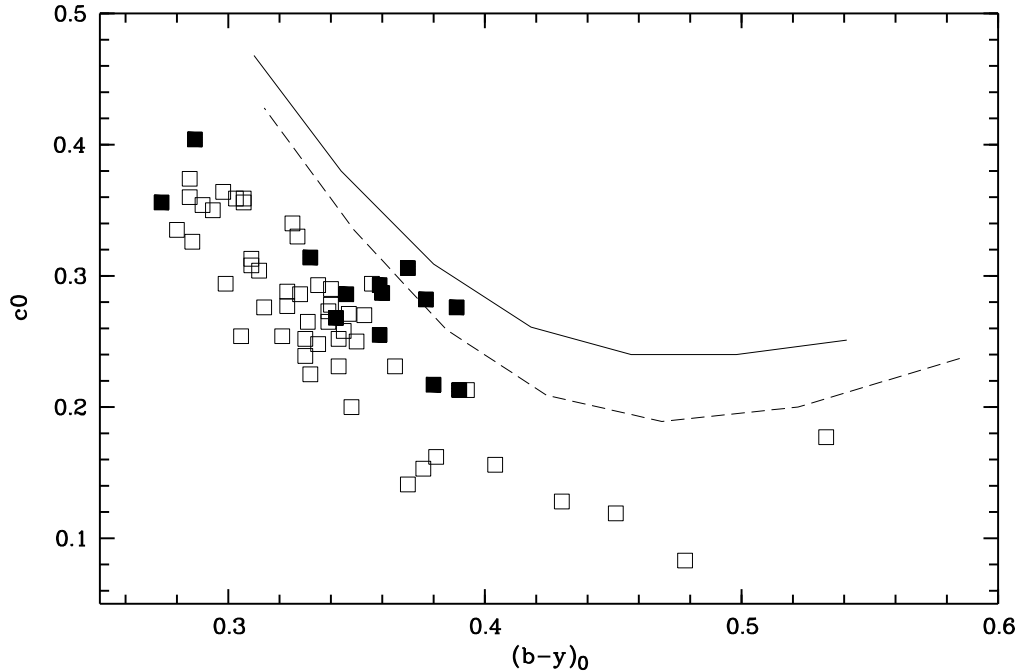}
\caption{ The $c0,(b-y)_0$ diagram for our sample of Pop II
stars. The solid line represents the locus of points
with log g = 3.5 for  [Fe/H]=-1.5. The dashed line
is the same but for  [Fe/H]=-3.0. The region above these lines
is populated by subgiants and giants.
 The filled squares are the stars with  [Fe/H] $> -1.5$
 while the open squares are stars with  [Fe/H] $\le -1.5$.
}
\end{figure}

 Abundances have been derived computing new atmospheric models 
by using the ATLAS9 code of Kurucz with enhanced 
$\alpha$-elements and
without the overshooting option. 
The theoretical EWs have been computed
by direct integration of synthetic spectra computed with the SYNTHE
code, 
thus taking into account the 
doublet structure of the Li 670.7 nm line.
The models used are important in  Li analysis, and systematic
differences among different authors of the order of 0.1 dex have been 
ascribed to different assumptions in 
the model computations (Molaro et al 1995) .

\section{ Results}

The results, 
based on 41 plateau stars with ${T}_{\rm eff}~$ $>$ 5700
and  [Fe/H] $\le-1.5$,  
give a  weighted mean
value 
${\rm  [Li] = 2.20\pm 0.016}$
 ([Li] =log(Li/H)+12).
In good agreement  with the value of  [Li]$=2.21\pm 0.013$
of Molaro et al (1995),
who used  spectroscopically derived temperatures.
 The standard error is compatible with the
estimated observational errors ($0.11\pm 0.03$ dex)
showing no
evidence for intrinsic dispersion. 
\begin{figure}
\plotone{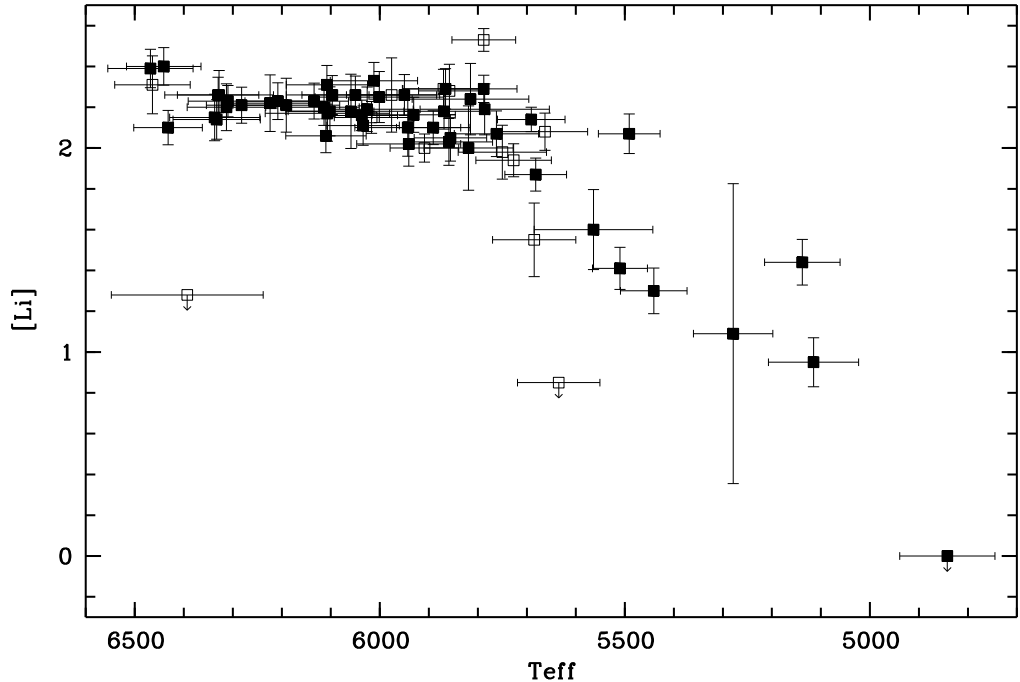}
 \caption{ The --  [Li] ${T}_{\rm eff}~$ diagram for our sample of stars.
 The filled squares are stars with  [Fe/H] $\le -1.5$ , while
 the open squares are those with  [Fe/H] $> -1.5$ .  Upper limits
 are shown with  downward arrows.
}
\end{figure}
No trends of  [Li] with  [Fe/H] are found  
over the range $-3.5\le \rm [Fe/H] < -1.5$.
A tiny trend with ${T}_{\rm eff}~$ is detected at 1 $\sigma$ level.
Our current best estimate, based on a bivariate fit is:
\begin{displaymath}
\rm { [Li]=1.146+0.033(\pm 0.046)
\times  [Fe/H]+0.018(\pm 0.011)\times
T_{eff}/100} 
\end{displaymath}
Our slope with ${T}_{\rm eff}~$ is smaller by about 
a factor of two than that found by
Norris et al (1994), Thorburn (1994) and Ryan et al (1996)
and is compatible with the mild depletion predicted in this
temperature range, by the standard models of Deliyannis et al (1990).

 The absence of a downturn in Li abundance at the hottest edge
of the plateau 
and the absence of dispersion on the plateau itself
argue strongly against significant depletion by diffusion 
or  rotational mixing.
(Vauclair 1988, Pinsonneault et al 1992).
The absence of a significant slope with ${T}_{\rm eff}~$ 
and the absence of intrinsic dispersion
rule out  stellar winds as possible  depletion mechanisms 
(Vauclair and Charbonnel 1995).  
All depletion mechanisms predict features that
are  present in the observed data.  Our results strongly support the 
view that the
observed  [Li] in Pop II stars coincides with the
primordial value.

\section{ Primordial Li}

A plausible   estimate for Li$_{p}$ can be deduced from
the weighted mean  of plateau stars  after the small trend
with T$_{eff}$, predicted by the standard models of
Deliyannis et al (1990), is accounted for;
thus leading to 
\begin{displaymath}
{\rm  [Li]_p = 2.240\pm 0.016_{1\sigma}\pm 0.05_{sys}}
\end{displaymath}
The systematic error follows from the error on the zero point
of the T$_{ eff}$ scale.
This corresponds to  
two possible values for $\eta = n_{B}/n_{\gamma}$. With  $\eta_{10} = 
10^{10}\eta$:
\begin{displaymath}
{ \eta_{10} 
= 1.7_{-0.3}^{+0.6}  ~~~~~~~~or~~~~~~~~ 
\eta_{10} =3.9 _{-1.0}^{+0.9} }
\end{displaymath}
But considering  the full errors 
both in the Li abundance and in the theoretical
BBN predictions the minimum of the 
Li SBBN prediction is allowed  and the two intervals merge.
Our low  $\eta$ is in agreement
with that 
obtained from  high deuterium values, D/H $\approx$ 10$^{-4}$ 
(Songaila et al 1994, Carswell et al
1996, Wampler et al 1996, Rugers and Hogan 1996) and with that obtained 
from primordial helium Yp =0.228 $\pm 0.005$ (Pagel et al 1992).
\begin{figure}
\plotone{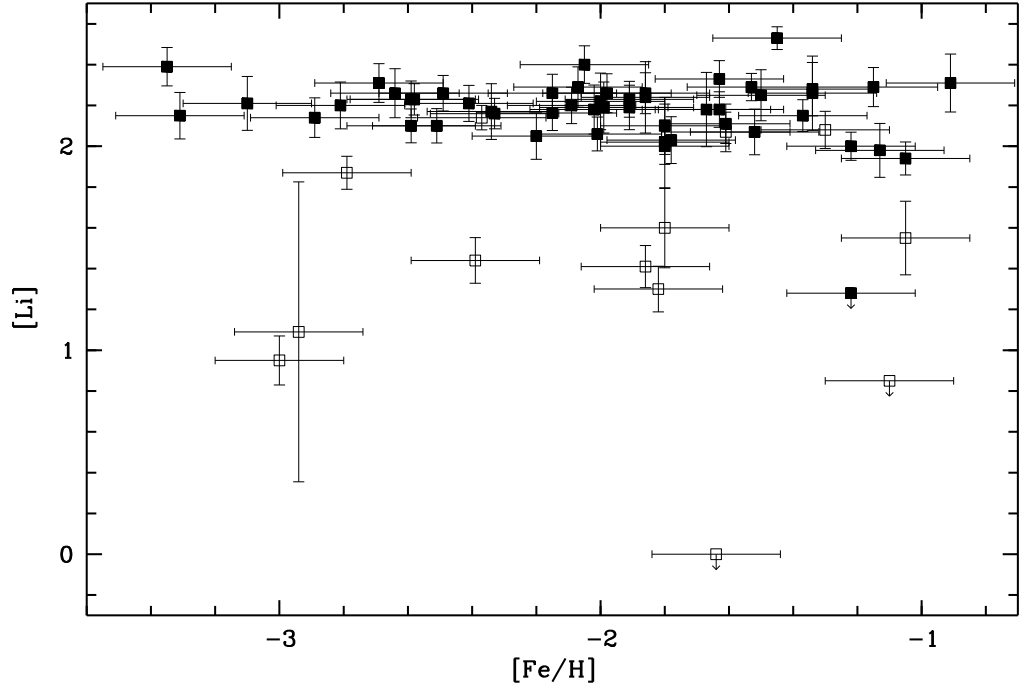}
 \caption{ The  [Li] --  [Fe/H] diagram for our sample of stars.
 The filled squares are stars with ${T}_{\rm eff}~$ $>$ 5700 K, while
 the open squares are those with ${T}_{\rm eff}~$ $\le$ 5700 K. 
 Upper limits are shown as downward arrows.
}
\end{figure}
 Our high  $\eta$  
 is more  consistent with the Yp=0.241 $\pm 0.003$
(Izotov et al 1994 ) 
who  use revised  neutral helium 
recombination coefficients, but it is 
remains inconsistent with the values from
the low D/H=2$\times 10^{-5}$ observed in high redshift absorption systems 
(Tytler Fan and Burles 1996, Burles and Tytler 1996 ). The consistency
between the low D/H value and Li requires a 0.5 dex of depletion
in Li which is not  supported by the present analysis. 
The two intercepts on the Li theoretical curve corresponds to two 
preferred  values for
the baryon density

\begin{displaymath}
{\rm \Omega_{B}h^2}  = 0.0063^{+0.0022}_{-0.0011} 
~~~~~~~~~or~~~~~~~~~{\rm \Omega_{B}h^2  = 0.0146^{+0.0033}_{-0.0037}} 
\end{displaymath}
where the Hubble parameter 
is H$_{0}$ = 100$h$ $\rm Kms^{-1}Mpc^{-1}$.
Considering that  $\Omega_{LUM}  
= 0.004+0.0007h ^{-3/2}$ (Persic and Salucci
1996), over the 
entire range of H$_{0}$ we have   $\Omega_{B} \ge \Omega_{LUM}$
suggesting the presence of dark baryons. 
The low value for $\eta$ when combined with the high
baryon fraction from X-ray of rich clusters leads to a total mass
$\Omega _{M} \le 0.2$, and to   the so called "X-ray Cluster Baryon
Catastrophe".

\end{document}